\documentclass[lettersize,journal]{IEEEtran}
\usepackage{amsmath,amsfonts}
\usepackage{algorithmic}
\usepackage{algorithm}
\usepackage{array}
\usepackage[caption=false,font=normalsize,labelfont=sf,textfont=sf]{subfig}
\usepackage{textcomp}
\usepackage{stfloats}
\usepackage{url}
\usepackage[utf8]{inputenc}
\usepackage{verbatim}
\usepackage{graphicx}
\usepackage{cite}
\usepackage{tikz}

\usepackage[colorlinks=true,linkcolor=blue,citecolor=blue]{hyperref}
\usepackage{verbatim}

\usepackage{colortbl}
\usepackage{xcolor}
\usepackage{array}
\usepackage{booktabs}
\definecolor{headerblue}{RGB}{102, 102, 153}
\definecolor{rowgray}{RGB}{238, 238, 238}

\usepackage{subcaption}
\makeatletter

\newcommand{\Rmnum}[1]{\expandafter\@slowromancap\romannumeral #1@}
\makeatother
\usepackage{balance}

\begin{document}
\title{Embodied Intelligent Spectrum Management: A New Paradigm for Dynamic Spectrum Access}

\author{Yihe Diao, Yuhang Wu, Hongtao Liang, Ming Xu, Rui Ding, Fuhui Zhou, \emph{Senior Member}, \emph{IEEE},\\ Qihui Wu, \emph{Fellow}, \emph{IEEE}, and Jun Zhang, \emph{Fellow}, \emph{IEEE}




\thanks{Y. Diao, H. Liang, M. Xu, R. Ding and Q. Wu are with the College of Electronic and Information Engineering, Nanjing University of Aeronautics and Astronautics, Nanjing, 210000, China (e-mail: diaoyihe@nuaa.edu.cn; ceie.lht@nuaa.edu.cn; xuming98@nuaa.edu.cn; rui\_ding@nuaa.edu.cn; wuqihui2014@sina.com).

Y. Wu and F. Zhou are with the College of Artificial Intelligence, Nanjing University of Aeronautics and Astronautics, Nanjing, 210000, China (e-mail:  may\_wyh@nuaa.edu.cn; zhoufuhui@ieee.org).

J. Zhang is with the Department of Electronic and Computer Engineering, The Hong Kong University of Science and Technology, Hong Kong, 999077, China (e-mail: eejzhang@ust.hk).}
}

\maketitle

\begin{abstract}

Wireless communication is evolving into an agent era, where numerous intelligent agents equipped with perception, reasoning, and interaction capabilities will operate in highly dynamic wireless environments. 
To complete diverse complex tasks, agent communication will play a critical role, which enables autonomous information exchange with external tools, services, and other agents.
This trendy movement will dramatically increase spectrum demand and result in unprecedented challenges for spectrum management. 
However, current spectrum management paradigms, including static spectrum allocation and intelligent management, lack the flexibility and generalization to accommodate the dynamic and heterogeneous demands of agent communication. 
The recent advancements in embodied intelligence (EI) bring a promising solution, and this article will provide our vision of an emerging embodied intelligent spectrum management (EISM) paradigm. 
We start with an architecture for EISM, elaborating its key enabling technologies. 
Then, a prototype platform is presented to demonstrate the advantages of EISM. 
Finally, key challenges and open issues are outlined to facilitate future research in this emerging field.

\end{abstract}

\begin{IEEEkeywords}
Agent communication, embodied intelligence, spectrum management, spectrum access, prototype platform
\end{IEEEkeywords}

\section{\scshape Introduction}

\IEEEPARstart{W}{ireless} communication is evolving into an agent era, where numerous intelligent agents (e.g., robotic agents, self-driving cars) equipped with perception, reasoning, and interaction capabilities are deployed in highly dynamic wireless environments. 
These agents rely on communication with various entities, including external tools, services, and other agents, and they interact with environments to complete diverse tasks.
Unlike traditional communication that primarily focuses on passive data transfer, agent communication is characterized by dynamic and task-driven information exchange \cite{agent_era}.
Consequently, this evolution drives a rapid increase in spectrum demand and brings unprecedented spectrum management challenges.
However, currently no dedicated spectrum bands have been specifically allocated for agent communication.
As the scale of agent deployment expands, the competition for limited spectrum resources intensifies, exacerbating severe spectrum interference and congestion, which significantly degrades communication reliability and hinders the agents from completing their tasks efficiently. 
Therefore, effective spectrum management becomes crucial for improving resource utilization and enabling reliable agent communication.

Currently, spectrum management predominantly relies on static allocation strategies that are fundamentally model-driven and depend heavily on extensive expert knowledge and predetermined rules \cite{static}. 
Under this paradigm, spectrum is pre-assigned to users or services by the administration. 
However, spectrum occupancy measurement campaigns conducted worldwide reveal that the spectrum utilization varies widely across bands, with several licensed bands exhibiting utilization rates below 15\% \cite{fcc}. 
In response, recent research has introduced intelligent spectrum management paradigms based on machine learning (ML), enabling dynamic, data-driven spectrum allocation that adapts to varying operational requirements and interference conditions \cite{ML_based}. 
However, the ML-based approaches are typically task-specific, exhibiting limited generalization across diverse scenarios and poor robustness in rapidly changing environments \cite{liu_llm}. 
Therefore, as wireless communication has to support massive deployments of agents operating in highly dynamic and decentralized environments, traditional spectrum management paradigms struggle to accommodate the task-driven and time-varying characteristics of agent communication.
\begin{figure*}[h]
    \centering
    \includegraphics[width=0.945\linewidth, trim=18 18 18 18,clip]{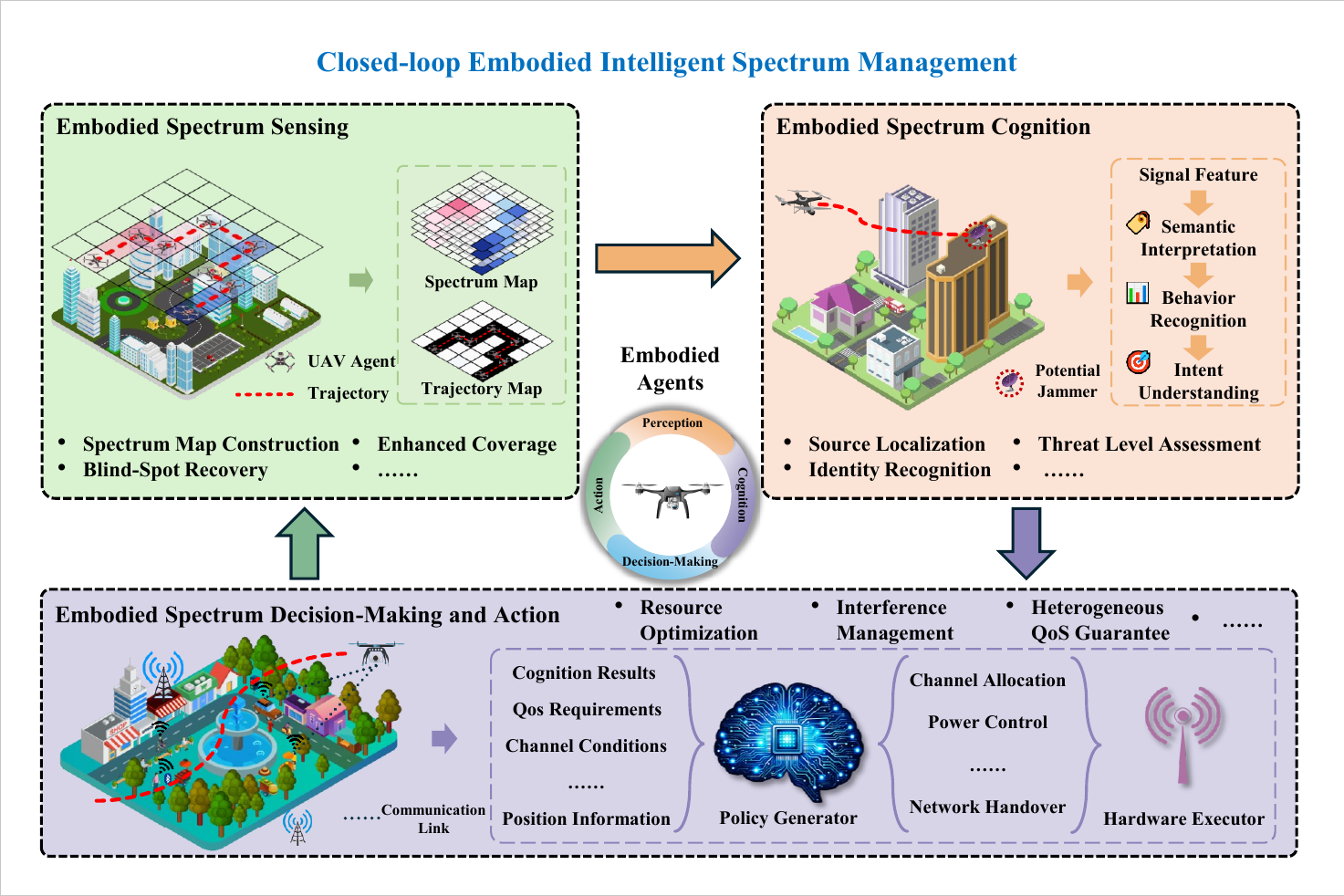}
    \caption{The architecture of the closed-loop embodied intelligent spectrum management.}
    \label{Fig.1}
    \vspace{-0.3cm} 
\end{figure*} 

The recent developments in embodied intelligence (EI) make it a promising solution for spectrum management. 
In particular, whereas traditional artificial intelligence (AI) technologies prioritize passive computational reasoning, EI emphasizes that intelligent behavior arises through active interaction between agents and their physical environment.
Through the dynamic loop among perception, cognition, decision-making and action, embodied agents can achieve real-time adaptation to environmental changes while actively shaping and transforming the environment \cite{embodied_intelligence}. 
This motivates us to propose embodied intelligent spectrum management (EISM) to exploit the advantages of embodied agents, to elevate spectrum management from passive response to active adaptation.
Specifically, EISM leverages embodied agents to autonomously sense spectrum conditions, reason about the spectrum demands of users, and coordinate spectrum access through the perception-action loop. 
Recent advances in foundation models, particularly large language models (LLMs), have shown promise for further enhancing the reasoning and generalization capabilities of embodied agents \cite{liu_llm}, making EISM particularly well suited for the complex and heterogeneous agent communication scenarios.
However, research on EISM remains nascent, with limited theoretical foundations and practical implementations.
This paper aims to explore and promote the research on EISM, with the goal of providing valuable insights and inspiration for this emerging field.

The main contribution of this paper lies in providing a comprehensive overview of the architecture of EISM and the key enabling technologies.
First, we present the architecture of EISM, including embodied spectrum sensing, cognition, decision-making, and action. 
We further discuss the key enabling technologies for EISM, and introduce a prototype platform to demonstrate the advantages of EISM.
Furthermore, the paper discusses several open issues and challenges in the process of EISM. 


\section{\scshape Architecture of Embodied Intelligent Spectrum Management}

EISM represents a promising direction for future spectrum management.
This section outlines its architecture, which comprises embodied spectrum sensing, cognition, decision-making, and action that form a closed-loop spectrum management paradigm, as illustrated in Fig. \ref{Fig.1} and compared with traditional intelligent spectrum management in Table \ref{tab:comparison}.

\subsection{Embodied Spectrum Sensing}

Spectrum sensing aims to determine whether a spectrum band is used by collecting and analyzing the radio signal.
Traditional spectrum sensing is typically achieved by using fixed sensing nodes or mobile vehicle-mounted sensing nodes, where the sensed spectrum band is fixed.
Although multiple mobile sensing nodes can improve the coverage flexibility \cite{ISAC_survey}, they typically operate with predefined trajectories and passive data collection, unable to actively explore and adapt to the evolving electromagnetic environment in real-time.
Moreover, traditional sensing approaches require high initial setup costs and maintenance expenses, thereby exhibiting limited flexibility across heterogeneous scenarios. 
As a result, blind spots and coverage gaps typically arise, particularly in complex urban environments, hindering comprehensive spectrum sensing.

To address these challenges, embodied spectrum sensing leverages the active environmental interaction capabilities of embodied agents to realize autonomous and adaptive sensing. 
Specifically, through continuous spatial exploration, embodied agents can progressively construct and refine spectrum maps that characterize the electromagnetic environment, while simultaneously generating trajectory maps that record effective sensing paths for future reuse.
This autonomous and adaptive sensing approach transforms spectrum sensing from static observation to active exploration, thereby potentially improving the comprehensiveness and accuracy of spectrum situational awareness.
For example, upon detecting coverage blind spots, an embodied unmanned aerial vehicle (UAV) agent can actively replan its trajectories to approach target areas for detailed investigation and blind-spot recovery.
Moreover, when signal attenuation is detected due to physical obstacles, the embodied UAV agent can autonomously navigate toward elevated positions to establish line-of-sight (LoS) communication links that offer more stable channel conditions, thereby further enhancing coverage and eliminating sensing dead zones.

\begin{table*}[h]
\centering
\normalsize
\caption{Comparison Between Traditional Intelligent Spectrum Management and Embodied Intelligent Spectrum Management}
\label{tab:comparison}
\renewcommand{\arraystretch}{1.2}
\setlength{\arrayrulewidth}{0.8pt}
\normalsize
\begin{tabular}{>{\centering\arraybackslash}m{2.46cm}|m{7.1cm}|m{7.1cm}}
\hline
\rowcolor{headerblue}
\rule{0pt}{1.3em}\textcolor{white}{\textbf{Module}} & \rule{0pt}{1.3em}\textcolor{white}{\textbf{Traditional Intelligent Spectrum Management}} & \rule{0pt}{1.3em}\textcolor{white}{\textbf{Embodied Intelligent Spectrum Management}} \\[0.3em]
\hline
\rowcolor{rowgray}
Sensing &  Passive data collection from fixed or mobile sensing nodes with predetermined trajectories. & Active exploration with autonomous trajectory replanning and blind-spot recovery. \\
\hline
Cognition &  Passive analysis of collected measurement data to extract signal features, highly dependent on data quality and completeness. & Interactive paradigm with a progressive process, enabling predictive and causal electromagnetic situational awareness. \\
\hline
\rowcolor{rowgray}
Decision-Making \& Action & Reactive response, passively adjusting communication parameters to environmental variations. &  Proactive adaptation with deep understanding of environmental dynamics, reasoning about consequences before physical actions. \\
\hline
\end{tabular}
\vspace{-0.3cm} 
\end{table*}



\subsection{Embodied Spectrum Cognition}

Spectrum cognition refers to the process of understanding and interpreting the electromagnetic environment.
Traditional spectrum cognition typically relies on passive analysis of collected measurement data to extract signal features.
However, such passive approaches are highly dependent on the quality and completeness of collected measurement data.
When measurements are incomplete, corrupted by severe noise, or captured from limited observation angles, cognitive accuracy degrades significantly, thereby hindering comprehensive understanding of the electromagnetic environment \cite{liu_llm}. 
These limitations are exacerbated in highly dynamic agent communication settings, where spectrum usage and interference evolve rapidly over space and time.

Embodied spectrum cognition overcomes these limitations by shifting to an interactive paradigm, in which embodied agents actively explore the environment to obtain comprehensive measurements and a deeper understanding of the electromagnetic environment.
In particular, embodied spectrum cognition follows a progressive process.
At the signal level, embodied agents extract fundamental signal features from measurement data collected at multiple spatial positions.
Building upon these features, semantic interpretation is performed by embodied agents to identify communication protocols and modulation schemes of the detected signals.
Further analysis enables behavior recognition, where embodied agents characterize the operational patterns and temporal characteristics of the emitters.
At the highest level, intent understanding integrates all preceding cognitive results to infer the underlying purpose and future actions of the detected emitters.
For instance, the embodied agents can actively approach a potential jammer to achieve more accurate source localization, conduct identity recognition by analyzing close-range signal characteristics, and ultimately assess the threat level based on the intent understanding, enabling a predictive and causal electromagnetic situational awareness. 

\subsection{Embodied Decision-Making and Action for Spectrum Management}

Spectrum decision-making and action form the final stage of spectrum management, processing observations and cognitive results into detailed management strategies and execution.
Conventional approaches typically operate in a reactive manner, passively responding to environmental variations by adjusting communication parameters such as operating spectrum band, transmit power, and modulation \cite{ding2025}.
However, such passive response is inherently constrained by the prevailing environmental conditions. 
Specifically, when severe interference or unfavorable propagation occurs, the optimization space becomes significantly limited, resulting in degraded system performance.
These challenges become more severe in task-driven agent communication scenarios, where spectrum conditions evolve rapidly and unpredictably.


To address these challenges, embodied spectrum decision-making and action is achieved by enabling embodied agents to actively understand and adapt to the environment rather than passively responding to it. 
With a deep understanding of environmental dynamics and physical constraints, embodied agents can reason about the consequences of candidate policies before physical actions, enabling more informed and effective decision-making. 
Furthermore, in order to effectively coordinate policies with physical actions, embodied spectrum decision-making and action exploit a hierarchical architecture comprising a policy generator and a hardware executor \cite{LM_decide}. 
Specifically, the policy generator integrates multi-dimensional information, such as cognition results, quality of service (QoS) requirements, channel conditions, and position information, to jointly optimize communication performance and physical movements. 
Moreover, by comprehensively understanding the interplay between spectrum conditions and physical movements, the policy generator can evaluate long-horizon outcomes and select actions that maximize expected performance while avoiding potential risks. 
The hardware executor then translates these policies into concrete operations, such as channel allocation, power control, network handover, and trajectory adjustment.

\section{\scshape Key Enabling Technologies}

\begin{figure}[!t]
    \centering
    \includegraphics[width=0.96\linewidth, trim=15 15 15 15,clip]{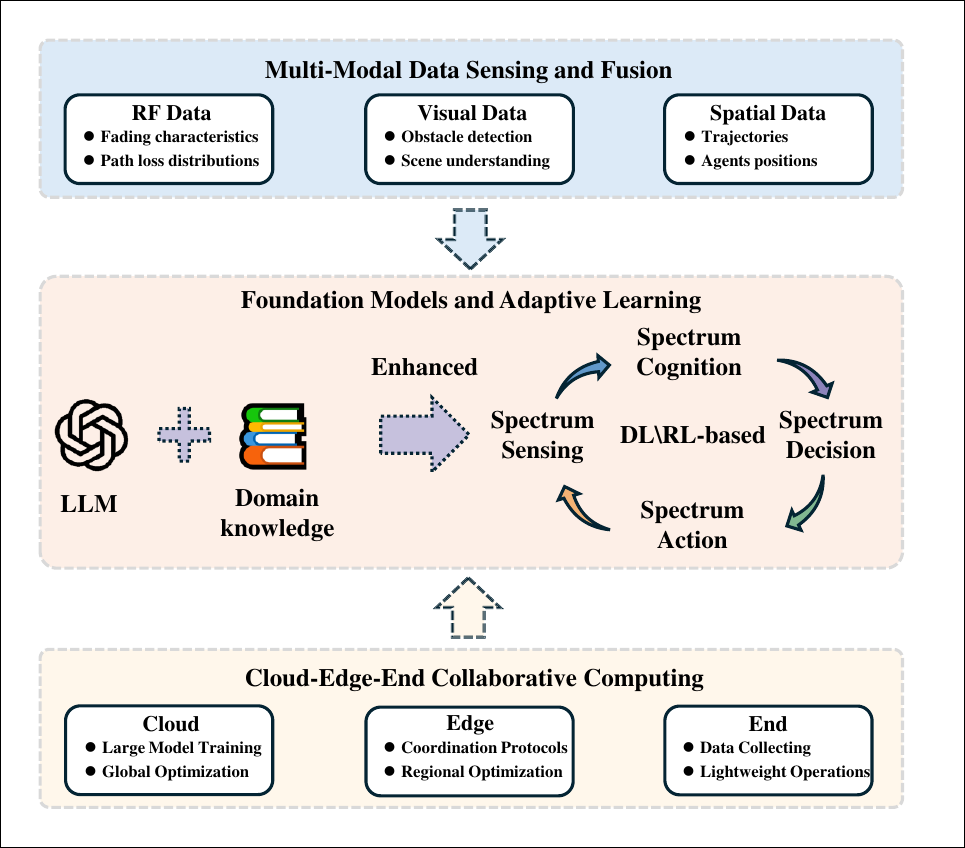}
    \caption{Technological pillars enabling EISM.}
    \label{Fig.2}
    \vspace{-0.3cm} 
\end{figure}

The realization of EISM relies on a series of advanced technologies, among which AI technologies play a particularly important role.
The key technologies are introduced in this section as pillars that mutually support and collectively enable EISM, as summarized in Fig. \ref{Fig.2}.

\subsection{Multi-Modal Data Sensing and Fusion}

Spectrum management systems inherently rely on accurate and comprehensive data to support effective decision-making.
Traditional spectrum management paradigms primarily depend on radio frequency (RF) signal measurements, such as received signal strength and spectral occupancy. 
However, this RF-centric approach overlooks the relationship between electromagnetic features and the physical environment, including terrain features, building distributions, and obstacle locations that significantly influence signal propagation characteristics. 
As a result, traditional approaches struggle to capture the complex spatial-temporal dynamics of the electromagnetic environment, limiting their adaptability in heterogeneous scenarios. 
Therefore, to enable EISM to achieve comprehensive situational awareness, real-time sensing and fusion of multi-modal data reflecting both electromagnetic features and the physical environment become essential.

Multi-modal data for EISM typically encompasses RF measurements, visual observations, and spatial information.
For example, the SynthSoM dataset \cite{synthsom} is a synthetic multi-modal sensing–communication dataset that provides a critical data foundation for EISM. 
In particular, it contains not only RF communication data (e.g., large-scale and small-scale fading characteristics and path-loss distributions) but also rich non-RF sensing data.
Furthermore, to capture the complex relationships between electromagnetic situation and physical environments, effectively fusing and processing these heterogeneous multi-modal data is essential for EISM.
For instance, multi-modal intelligent channel modeling (MMICM) techniques can enable embodied agents to intelligently process heterogeneous information streams from communication devices and multi-modal sensors, thereby capturing the complex and dynamic relationships across sensing modalities \cite{mmicm}. 
Through advanced and effective data sensing and fusion, EISM can construct unified representations that encode both signal distributions and environmental context, thereby transforming raw multi-modal observations into physical actions for adaptive spectrum management.

\subsection{Foundation Models and Adaptive Learning}

As wireless communication evolves into the agent era, agent communication exhibits inherent task-driven characteristics that introduce unprecedented dynamics into spectrum management. 
In particular, the spectrum demands of agent communication fluctuate rapidly based on heterogeneous task requirements, environmental interactions, and real-time coordination needs. 
To enable EISM to cope with such highly dynamic electromagnetic environments, adaptive learning mechanisms are essential, allowing embodied agents to continuously adjust their spectrum management strategies through interaction with their environment.


Deep learning (DL) and reinforcement learning (RL) have made significant contributions to enable the adaptive learning capabilities of embodied agents.
For example, DL-based approaches excel at pattern recognition tasks including signal classification, modulation identification, and interference detection, achieving high accuracy through data-driven feature learning \cite{liu_llm}. 
Meanwhile, RL can enable embodied agents to learn spectrum management strategies and trajectory optimization end-to-end through trial-and-error interaction with the environment.
For instance, a deep reinforcement learning (DRL) framework combined with active inference has been investigated to address joint spectrum allocation and trajectory optimization in partially observable environments by incorporating demand prediction and multi-agent collaboration \cite{ding2025}. 
Furthermore, advanced training approaches such as digital cousin technologies enable parallel policy learning across multiple virtual environments, significantly accelerating the convergence and enhancing the robustness of policies for large-scale deployments \cite{digital_cousin}. 
Despite these advantages, existing DL-based and RL-based approaches exhibit fundamental limitations when applied to EISM. 
Specifically, these approaches typically operate as specialized modules for individual tasks, lacking the unified reasoning capability to jointly process observations, domain knowledge, and objectives across sensing, cognition, and decision-making processes simultaneously. 
As a result, these approaches struggle to maintain coherent and consistent performance when deployed in heterogeneous agent communication scenarios.

Foundation models, particularly LLMs, are promising for further enhancing the adaptive learning capabilities of EISM by simultaneously leveraging the semantic understanding, domain knowledge, and cross-task reasoning. 
In particular, unlike task-specific DL models, LLMs can leverage pre-trained knowledge about electromagnetic principles, communication protocols, and spectrum regulations, enabling flexible adaptation to new scenarios without extensive retraining.
For instance, SpectrumFM combines spectrum knowledge graphs with empirical data to address critical challenges such as weak signal detection, few-shot recognition, and the identification of unknown devices in open-set environments, thereby enabling robust signal classification and intent inference even with limited training samples \cite{liu_llm}. 
For embodied spectrum decision-making, LLMs can generate contextually appropriate strategies by reasoning over environmental conditions and operational constraints. 
Recent work has further demonstrated the effectiveness of LLM-enhanced DRL approaches for resource allocation in wireless communication, where LLMs serve as high-level cognitive cores to provide semantic guidance and domain knowledge, while DRL handles low-level control and continuous policy optimization \cite{llm_uav_opt}. 
This synergistic architecture can enable embodied agents to achieve both generalization across diverse spectrum scenarios and continuous adaptation to dynamic electromagnetic environments.

\subsection{Cloud-Edge-End Collaborative Computing}

EISM leverages numerous advanced AI technologies to enable intelligent spectrum management.
However, these capabilities fundamentally depend on processing massive amounts of multi-modal data through computationally intensive algorithms \cite{edge_computing}. 
The substantial computational demands pose significant challenges for deployment on resource-constrained embodied platforms such as UAVs and unmanned ground vehicles (UGVs), potentially degrading real-time spectrum sensing and decision-making performance.

A hierarchical cloud-edge-end collaborative computing architecture promises to address these challenges by distributing EISM tasks according to their latency requirements and computational complexity \cite{edge_framework}.
Specifically, at the end layer, embodied agents execute lightweight operations including real-time spectrum sampling, local signal detection, immediate spectrum access actions, and basic signal processing tasks that require minimal latency. 
The edge layer handles region-specific spectrum management tasks, including local spectrum map construction, interference source identification and localization, and coordination protocols among nearby agents.
Meanwhile, the cloud layer serves as the computational backbone for training spectrum foundation models, conducting global spectrum optimization, maintaining comprehensive spectrum knowledge bases, and formulating long-term spectrum allocation strategies.
This hierarchical architecture ensures that latency-sensitive spectrum sensing and access operations are executed locally, while computation-intensive cognition and global optimization tasks are offloaded to more powerful infrastructure, thereby enabling EISM to achieve both real-time responsiveness and system-wide intelligence.

\section{\scshape Prototype Platforms}
Building upon the proposed architecture of EISM, we detail the design and validation of a hybrid EISM platform in this section. 
The platform operationalizes the EI paradigm by implementing a closed-loop system where embodied agents collaborate with ground static infrastructure to achieve continuous spectrum sensing, cognition, decision-making, and action.

\subsection{Overall Architecture of the EISM Platform}

The platform adopts a hierarchical hybrid architecture designed to organically integrate the persistence of static sensing infrastructure, i.e., fixed monitoring nodes, with the mobility of embodied agents, i.e., UGVs and UAVs, thereby achieving robust, efficient, and adaptive spectrum management.
Specifically, the architecture is composed of three tightly coupled layers: a static sensing layer, an EI layer, and a centralized intelligence and control layer.

\begin{figure}[!t]
    \centering
    \includegraphics[width=0.94\linewidth]{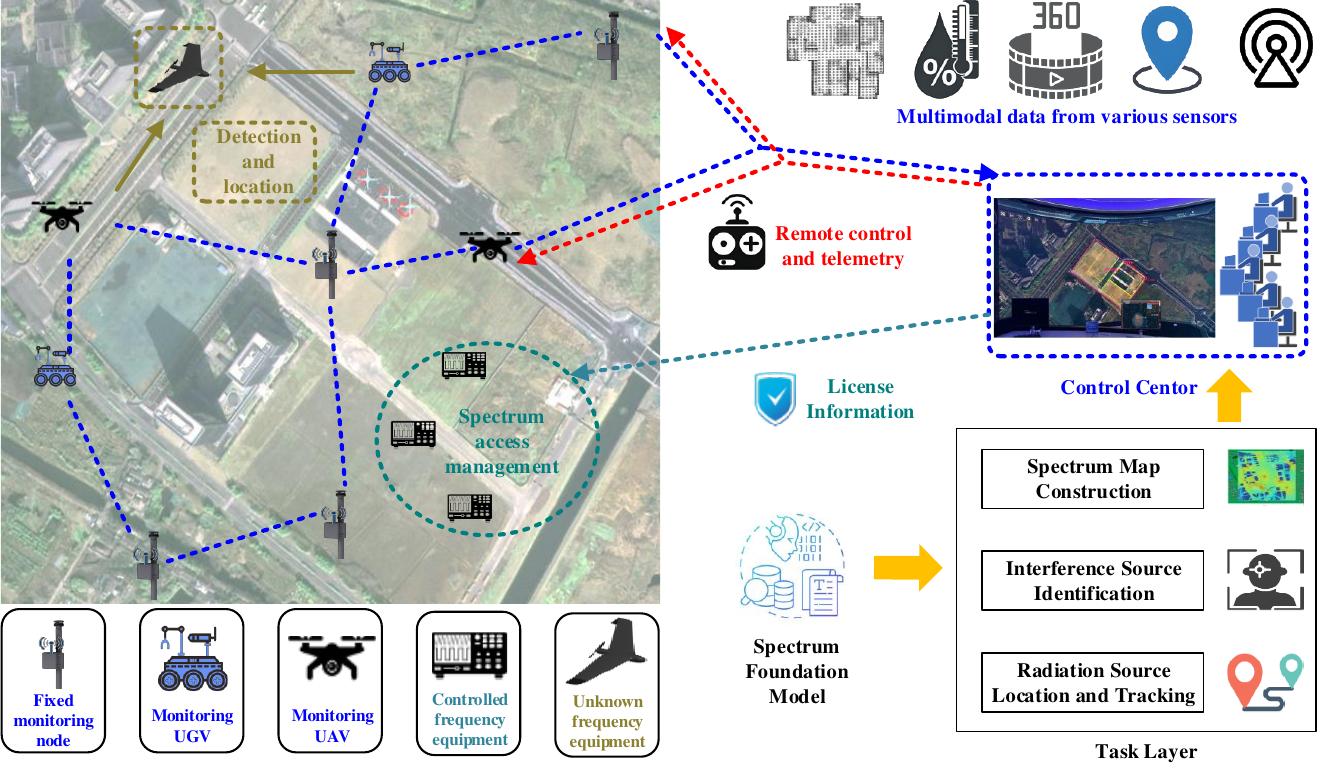}
    \caption{An EISM platform for the Nanjing Low-Altitude Integrated Flight Base.}
    \label{fig:platform}
     \vspace{-0.3cm} 
\end{figure}

The static sensing layer serves as the foundation for routine monitoring, consisting of widely deployed ground-based fixed spectrum monitoring nodes.
These nodes are equipped with various sensors that collect multi-modal data, including temperature, humidity, pressure, panoramic video, ranging and positioning data, and RF data.
Moreover, these nodes are responsible for conducting extensive continuous spectrum scanning, establishing a historical baseline of spectrum occupancy, and performing initial detection of anomalous interference signals. 
Through internal collaboration among nodes, coarse-grained localization of anomaly sources can be achieved.
As a result, the static sensing layer can provide wide-area, persistent situational awareness and form the first line of defense and data foundation for the EISM platform.

The EI layer acts as an intelligent and agile supplement to the EISM platform, comprising inspection UAVs and UGVs equipped with spectrum sensors and computing units. 
In contrast to the continuous operation mode of the static nodes, the embodied agents are not perpetually active but are instead triggered and dispatched by the central system. 
When anomalies are detected by the static sensing network but cannot be accurately localized, or when detailed inspection of a specific region is required, the embodied agents are deployed to the designated area.
By exploiting their unique mobility, they perform high-precision, targeted tasks that the static network cannot perform, such as close-range fine-grained measurement of suspicious signals, precise traceability and localization, and executing active adaptations like dynamic spectrum access.
Through these capabilities, the EI layer can extend the perception and action scope of the system beyond the coverage limitations of static sensing infrastructure.

The centralized intelligence and control layer functions as the brain of the entire EISM platform, responsible for coordinating the resources of both the static and dynamic layers to form closed-loop decision-making. 
In particular, this layer incorporates a spectrum foundation model, which is continuously updated with multi-source data from both the fixed monitoring nodes and the embodied agents.
For example, the powerful embodied spectrum cognition and decision-making core constructs a spectrum map that supports interference and radiation source analysis and high-level decision-making, including agent dispatch and mission planning.
Finally, a centralized orchestrator processes the decisions into concrete commands, scheduling the mission execution of embodied agents and managing the spectrum access of controlled frequency equipment.
Through the collaboration of these three layers, the platform establishes a complete closed loop from wide-area spectrum sensing to precise spectrum action, embodying the core concept of EISM.

\subsection{Case Study: Embodied Spectrum Mapping in Urban Environment}
\begin{figure}[!t]
    \centering
    \includegraphics[width=0.97\linewidth,trim=15 15 15 15,clip]{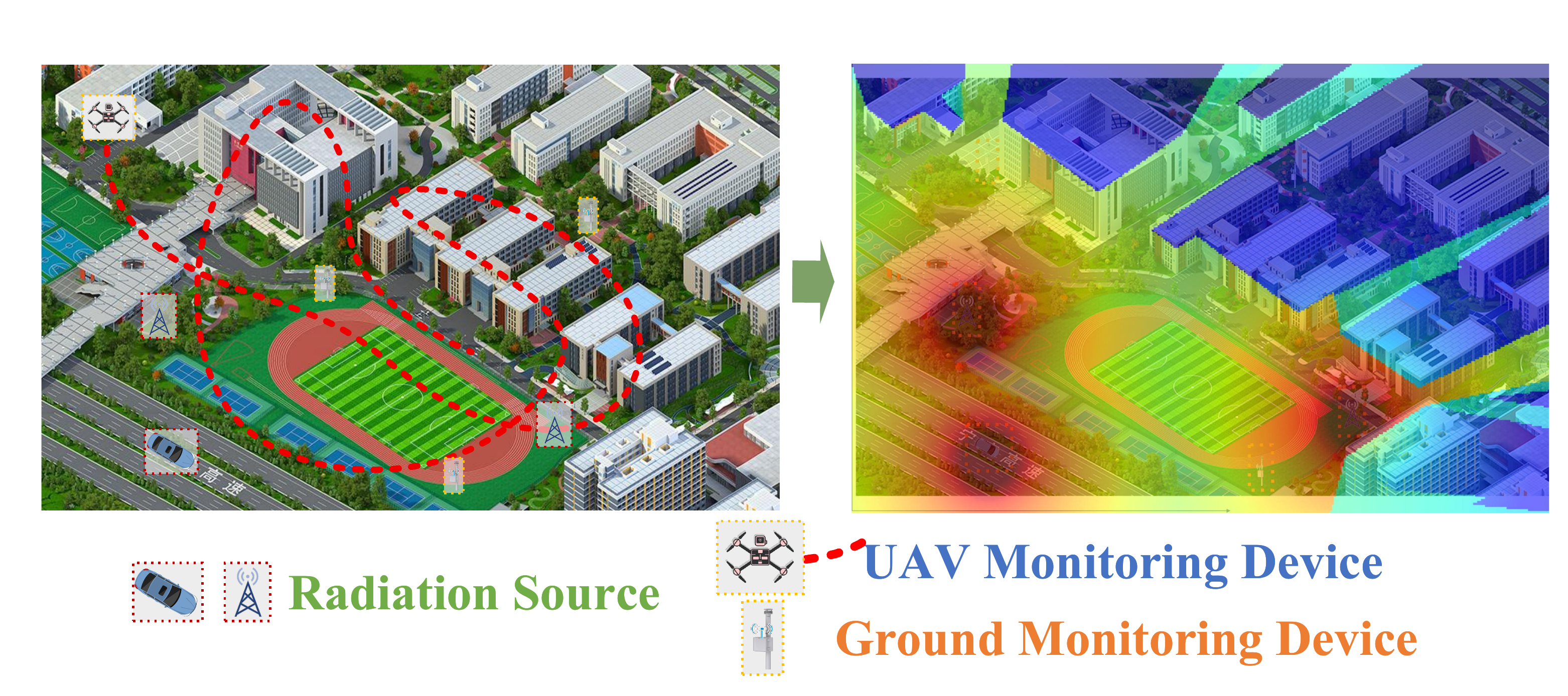}
    \caption{Embodied agents collaboratively construct a 3-D urban spectrum map through coordinated active sensing.}
    \label{fig:EmbodiedSpectrum}
    \vspace{-0.3cm} 
\end{figure}

High-fidelity three-dimensional (3-D) spectrum maps are a foundational element for future intelligent wireless communications. 
Traditional construction approaches suffer from fundamental limitations due to their reliance on data collected by fixed sensors.
As a result, when propagation conditions become highly heterogeneous, particularly in complex urban environments, propagation effects like occlusion and reflection from buildings create extensive information islands and sensing blind spots, making it impossible to form global and continuous spectrum situational awareness.
To address these challenges, our case study validates a novel solution based on the EISM platform, which leverages multiple types of embodied agents to build deep cognitive awareness about their physical environments through coordinated active exploration.
Specifically, the solution integrates fixed monitoring nodes, UGVs, and UAVs, each operating as an embodied agent with sensing, cognition, and autonomous action capabilities.

Within this solution, fixed ground monitoring nodes provide persistent baseline observations of the electromagnetic environment and highlight regions with high uncertainty or abnormal signal patterns, forming an initial situational awareness.
Based on this initial awareness, UGVs autonomously navigate along road networks to investigate areas of interest, coupling each measurement with precise spatiotemporal coordinates while adapting their trajectories based on real-time findings. 
UAVs complement this ground-level exploration by accessing the 3-D spaces unreachable by ground vehicles, particularly capturing vertical signal distributions around building facades and rooftops.
All embodied agents process their collected data using probabilistic frameworks such as Gaussian Processes, which are used to construct a shared continuous cognitive map providing both predicted signal strength and uncertainty measures for every location. 
This uncertainty quantification enables each embodied agent to identify where collective understanding remains poor and to coordinate subsequent exploration strategies, allowing the multi-agent system in the EISM platform to autonomously maximize information gain and progressively refine the unified spectrum map.

The final constructed spectrum map visualized in Fig. \ref{fig:EmbodiedSpectrum} demonstrates the synergistic fusion of data from all nodes into a unified representation, overlaying the electromagnetic measurements onto a realistic 3-D urban model with colored peaks representing radio signal strength ranging from blue for weak signals to red for strong signals. The fixed nodes contribute temporal stability patterns, the ground vehicle trajectories expose street-level signal canyons and coverage gaps, and the UAV measurements complete the vertical dimension, revealing how signals propagate across building tops and through aerial corridors. This case study provides preliminary validation that coordinated active exploration by heterogeneous embodied agents can effectively address the coverage limitations inherent in traditional passive monitoring approaches. Moreover, the resulting high-fidelity spectrum map establishes a foundational cognitive layer for advanced tasks such as interference prediction and dynamic spectrum access, illustrating the potential of the EISM platform for addressing future spectrum management challenges.

\section{\scshape Challenges and Open Issues}
\begin{figure}[t]
    \centering
    \includegraphics[width=0.955\linewidth, trim=15 15 15 15,clip]{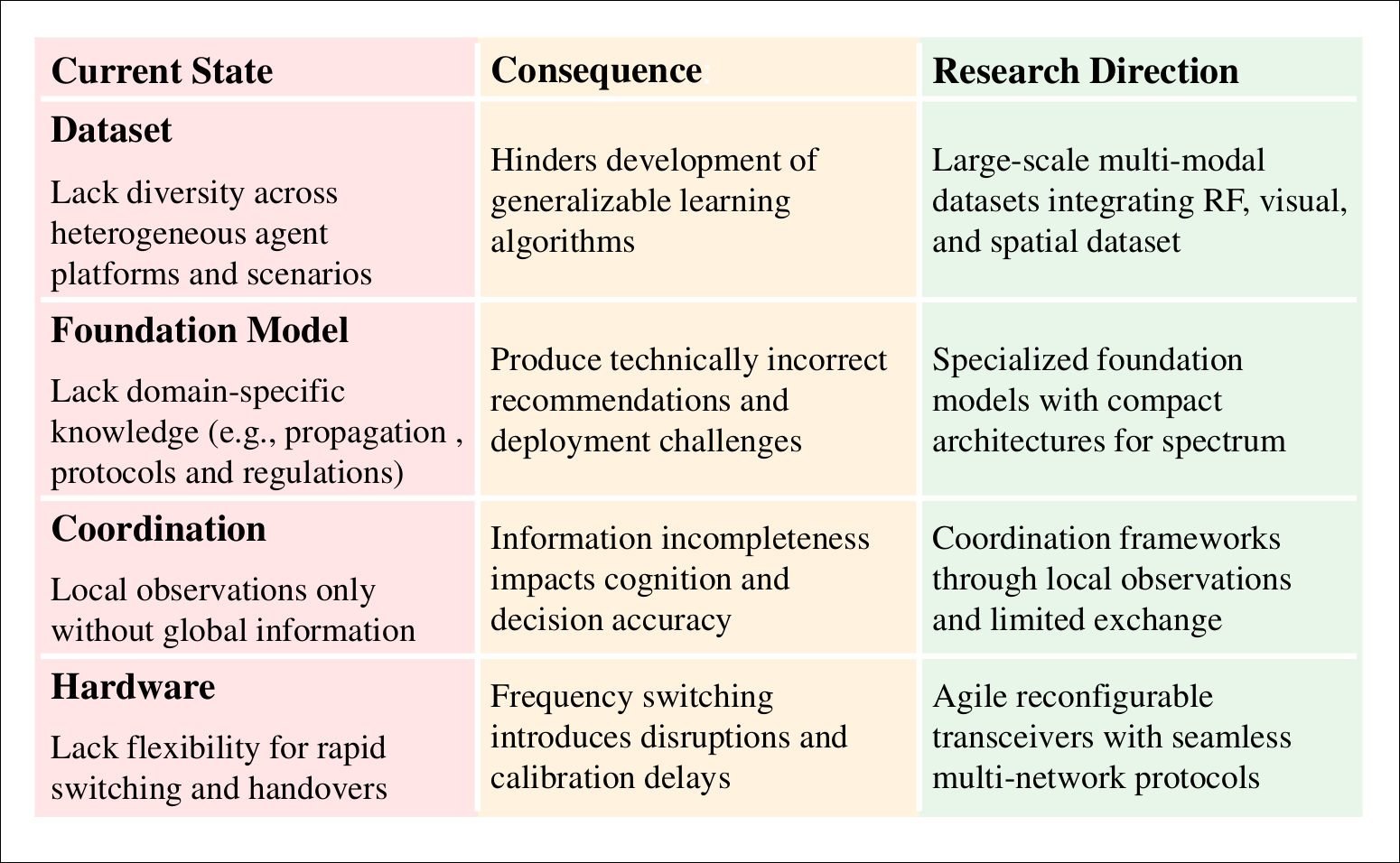}
    \caption{Key challenges and open issues in EISM.}
    \label{Fig.5}
    \vspace{-0.3cm} 
\end{figure}

EISM offers numerous advantages, including the potential to enhance adaptability and autonomy, enable widespread application scenarios, and provide powerful generalization capabilities. However, several core challenges must be addressed to realize its full potential in practical deployments. Fig. \ref{Fig.5} summarizes the key open issues associated with the realization of this paradigm.

\subsection{High-Quality Multi-Scenario Datasets for Heterogeneous Agents}

Current datasets lack the diversity required for EISM across heterogeneous embodied platforms and scenarios. 
Specifically, various embodied agents exhibit diverse physical characteristics, sensing capabilities, and mobility constraints, ranging from ground robots to aerial UAVs. 
Moreover, wireless environments vary significantly across urban, rural, indoor, and industrial settings, each with distinct propagation characteristics and interference patterns. 
The absence of comprehensive datasets encompassing these diverse combinations of agent communication scenarios hinders the development of generalizable learning algorithms.
Addressing this gap requires collaborative efforts to construct large-scale, multi-modal datasets that integrate RF measurements, visual observations, and spatial trajectories across diverse platforms and environments, thereby enabling robust cross-scenario generalization.

\subsection{Specialized Foundation Models for Spectrum Management}

General-purpose foundation models lack domain-specific knowledge critical for spectrum management, including electromagnetic propagation principles, communication protocols, and spectrum regulations. 
As a result, they often produce semantically plausible but technically incorrect recommendations due to insufficient understanding of physical layer characteristics. 
Moreover, their extensive parameter counts create deployment challenges for resource-constrained embodied platforms. 
Therefore, developing specialized foundation models pre-trained on electromagnetic spectrum knowledge graphs, communication standards, and propagation models is essential. 
These domain-specific models should maintain compact architectures suitable for edge deployment while generating technically accurate strategies. 
Hybrid approaches combining lightweight specialized models for domain reasoning with general models for high-level planning may offer practical solutions.

\subsection{Large-Scale Coordination Under Incomplete Information}

Scaling EISM to numerous distributed agents introduces fundamental coordination challenges. 
Individual agents possess only local observations and lack complete information about the global electromagnetic environment, the states of other agents, or network-wide spectrum utilization. 
This information incompleteness directly impacts spectrum cognition and decision-making accuracy, potentially causing misidentification of interference sources or suboptimal spectrum access decisions. Traditional centralized approaches become infeasible at scale, while fully decentralized methods often struggle to maintain consistency. Developing coordination frameworks that enable effective spectrum management through local observations, limited information exchange, and distributed decision-making while maintaining acceptable accuracy represents a critical research challenge requiring novel approaches in multi-agent learning and distributed optimization.

\subsection{Hardware Flexibility for Dynamic Spectrum Operations}

The effective operation of EISM demands hardware capable of rapid frequency switching, network handovers, and transmission parameter adjustments in response to dynamic conditions. 
Current RF front-ends and baseband processors often lack sufficient flexibility, with frequency switching introducing transient disruptions, calibration delays, and power consumption spikes.
Meanwhile, the network transitions between heterogeneous technologies require complex protocol adaptations that strain computational resources. 
Moreover, embodied platforms face stringent size, weight, and power constraints limiting the feasibility of highly flexible software-defined radio architectures.
Therefore, advancing hardware toward agile, low-latency, energy-efficient reconfigurable transceivers, along with developing lightweight protocols for seamless multi-network operation, is essential to enhance EISM in practice.

\section{\scshape Conclusion}

A novel and promising spectrum management paradigm, EISM, was proposed, which leverages the autonomous capabilities of embodied agents to interact with the electromagnetic environment.
The architecture of EISM was presented and its substantial potential in advancing the efficiency of spectrum management was highlighted.
Moreover, the key enabling technologies for EISM were outlined. 
Furthermore, several challenges remained in realizing the full potential of EISM. Finally, we highlighted these open issues as future research directions to encourage further exploration and innovation, ultimately advancing EISM toward practical deployment in next-generation wireless communication.

\end{document}